\documentclass[aps,prl,twocolumn,a4paper,groupedaddress,twocolumn,showkeys,showpacs,floatfix]{revtex4}
\usepackage{graphicx}
\usepackage{tabularx}
\usepackage{amsmath}
\usepackage{amsfonts}
\usepackage{amssymb}
\usepackage{natbib}
\begin{document}
\title{Quantitative Determination of Enhanced and Suppressed Transmission through Subwavelength Slit Arrays in Silver Films}
\date{\today}
\author{D. Pacifici}\author{H.~J. Lezec}\author{Harry A. Atwater}\affiliation{Thomas J. Watson Laboratories of Applied Physics, California Institute of Technology\\Pasadena, California 91125}
\author{J. Weiner}\affiliation{IRSAMC/LCAR\\
Universit\'e Paul Sabatier, 118 route de Narbonne,\\31062
Toulouse, France\\and\\IFSC/CePOF\\ Universidade de S\~ao Paulo, Avenida Trabalhador S\~ao-calense,
400-CEP 13566-590\\ S\~ao Carlos SP, Brazil}
\email{jweiner@irsamc.ups-tlse.fr}
\begin{abstract}
Measurement of the transmitted intensity from a coherent monomode light source through a series of subwavelength slit arrays in Ag films, with varying array pitch and number of slits, demonstrate enhancement (suppression) by as much as a factor of $6$ ($9$) when normalized to that of an isolated slit.  Pronounced minima in the transmitted intensity were observed at array pitches corresponding to $\lambda_{\mathrm{SPP}}$, $2\,\lambda_{\mathrm{SPP}}$ and $3\,\lambda_{\mathrm{SPP}}$ where $\lambda_{\mathrm{SPP}}$ is the wavelength of the surface plasmon polariton (SPP).  Increasing the number of slits to more than four does not increase appreciably the per-slit transmission intensity.  These results are consistent with a model for interference between SPPs and the incident wave that fits well the measured transmitted intensity profile.
\end{abstract}
\maketitle
Since the first experimental report of ``extraordinary optical transmission" through subwavelength hole arrays\,\cite{ELG98}, considerable effort has been devoted to the essential physics of the process in both hole\,\cite{PNE00,MGL01,CEW03,CGS05,GSC06} and slit\,\cite{PGP99,LHA00,T02,CL02,XZM05} arrays.  Some studies\,\cite{PNE00,MGL01,PGP99} have interpreted the transmission spectrum as excitation of delocalized surface plasmon Bloch modes and identified peaks in the transmission with wavelengths equal to integer multiples of the array pitch.  Others\,\cite{CL02,LSH03,XZM05,JUH06} concluded that this condition should be associated with transmission minima.  Still other theoretical work\,\cite{SVV03,CEW03,CGS05} interpreted the transmission line shapes as ``Fano profiles," involving interferences between surface and incident propagating modes.  Further spectral transmission measurements\,\cite{LT04} revealed that, normalized to a single aperture, suppression as well as enhancement was a characteristic property of hole and slit arrays, and interferometric studies\,\cite{GAV06a,GAV06b,KGA07} showed that the contribution of transient diffracted surface modes are as important as the surface plasmon polariton (SPP) guided mode in the immediate vicinity of the subwavelength object.  The experimental setup reported in Refs.\,\cite{ELG98,LT04} consisted of an incoherent, broad-band light source passed through a scanning spectrophotometer and focused on fixed-period subwavelength hole and slit arrays.  Transmitted intensity was detected in the far field as a function of the scanned wavelength.  In that work the spectral resolution and coherence length of light incident on the arrays therefore depended on instrumental parameters, and these in turn can affect the position and shape of the measured spectral features.  Furthermore, the frequency-dependence of the dielectric constant of Ag and other real metals is non-negligible in the range of typical wavelength scans from 450 to 900~nm.

\begin{figure}
  \includegraphics[width=0.8\columnwidth]{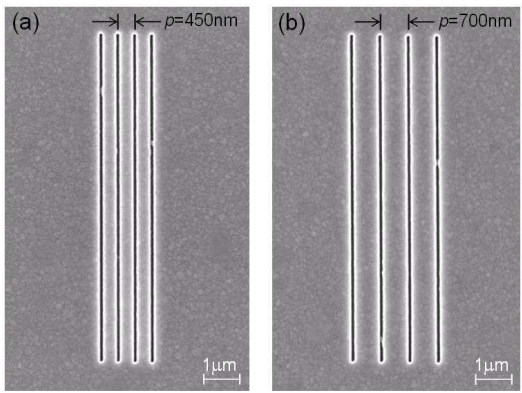}
  \caption{Two typical elements in the overall structure layout.  Panel (a) shows $N=4$, $p=450$~nm. Panel (b) shows $N=4$, $p=700$~nm.  Each slit is FIB milled through a 200~nm thick silver layer.  Dimensions of each slit are 50~nm wide and 10~$\mu$m long.}\label{Fig-SEM-image}
\end{figure}

\begin{figure}\centering
    \includegraphics[width=1.1\columnwidth]{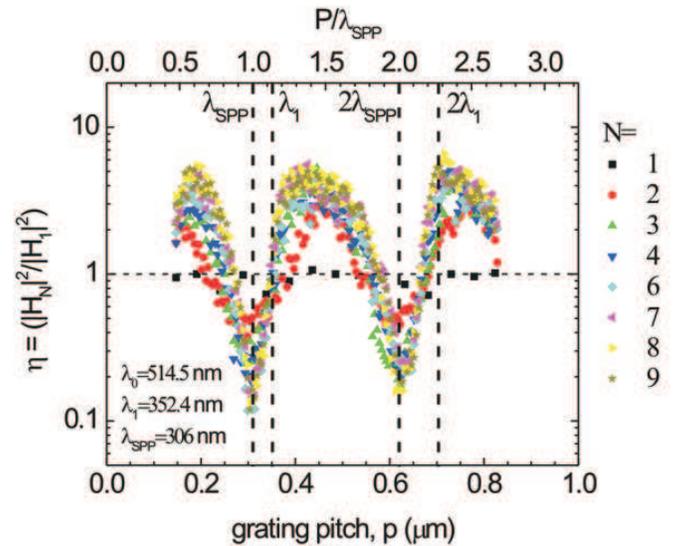}\caption{(color online) Normalized transmission intensity $\eta$ vs. grating pitch (in micrometers on the lower abscissa and normalized to $\lambda_{\mathrm{SPP}}$ on the upper abscissa) for a series of slit arrays $N=1-9$.  Gratings with $N=5$ were omitted due to defective fabrication.  The wavelengths $\lambda_0$, $\lambda_1$, and $\lambda_{\mathrm{SPP}}$ are, respectively, the free-space wavelength, the wavelength in fused silica ($n=1.46$) and the wavelength of the surface plasmon polariton.}\label{Fig.Intensity-vs-Pitch}
\end{figure}

In order to remove these measurement ambiguities and experimentally test the various theoretical interpretations, we have undertaken a series of high-resolution measurements of the transmission through slit arrays in which the spectral source is coherent, monomode and at fixed frequency.  The transmission measurement setup consists of a $\lambda=514.5$~nm, 5~mW, TEM$_{00}$ light beam from an Ar ion laser and aligned to the optical axis of an inverted microscope.  The beam is focused at normal incidence onto the sample surface through the microscope condenser and polarized TM (magnetic H-field component parallel to the long axis of the slits). Light intensity transmitted through each slit array is then gathered by a 50X microscope objective with a numerical aperture (NA) of 0.45 and detected with a liquid-nitrogen-cooled, charged-coupled device (CCD) array detector.  Light intensity transmitted through each slit array is obtained by integrating the signal over the entire region of interest in the CCD image and subtracting the background originating from electronic noise. Per-slit transmission intensities are obtained by correcting the transmitted intensity for the calculated collection efficiency of the microscope objective lens and normalizing the transmitted intensity for each series of gratings to the intensity collected from a single-slit structure.  The series of slit arrays were milled with a focused-ion-beam (Ga$^+$ ions, 30~keV) in a 200~nm thick layer of silver evaporated onto a flat fused-silica microscope slide.  The layout of slit-array structures consisted of a matrix of 9 rows and 140 columns.  Each row was indexed by the slit number $N$ in the array and varied from $N=1-9$.  Each column was indexed by the array pitch $p$ starting from the first column at $p=150$~nm and incremented by 5~nm with each successive column.  Thus the pitch varied over a range from $p=150-845$~nm; from less than $\lambda_{\mathrm{SPP}}$ to greater than $2\,\lambda_{\mathrm{SPP}}$.  Each slit was milled 50~nm wide, 200~nm deep, and 10~$\mu$m long as shown in Fig.\,\ref{Fig-SEM-image}.
The structured silver layer was covered by a second microscope slide; optically contacted to the silver surface by index-matching fluid ($n=1.46$) so that the index change at the dielectric-silver interface was identical at both the input (incident) and output (transmitted) planes.  The transmitted intensity of each successive array along a given row was recorded in the far-field by the CCD as the sample was stepped using an X-Y translation stage.  The results are summarized in Fig.\,\ref{Fig.Intensity-vs-Pitch}.

Taking into account the collection efficiency of the microscope objective and the far-field angular distribution of the slit grating diffraction modes, we define $\eta=|H_N|^2/|H_1|^2$ as the ratio of the magnetic field intensity at the output aperture of each slit in an array of $N$ slits to the magnetic field intensity at the output aperture of an isolated slit.  Figure \ref{Fig.Intensity-vs-Pitch} plots $\eta$ vs. array pitch for all arrays $N=1-9$ (except $N=5$, omitted due to defective fabrication).  The results show that the transmission intensity for all arrays exhibits very similar behavior with transmission dropping to a minimum $\simeq 0.1\,\eta$ at an array pitch equal to $\lambda_{\mathrm{SPP}}$, then rising to a broad maximum $\simeq 6\,\eta$ before repeating similar behavior around $2\,\lambda_{\mathrm{SPP}}$.  The position of the minima are in accord with earlier predictions\,\cite{CL02,LSH03} and simulations\,\cite{XZM05,XZM06} and at variance with others\,\cite{PNE00,MGL01} predicting transmission \emph{maxima} at $\lambda_{\mathrm{SPP}}$.

The wavelength $\lambda_{\mathrm{SPP}}$ was calculated from the usual formula for the guided surface wave\,\cite{Raether88},
\parbox{0.8\columnwidth}
{\begin{eqnarray*}
n_{\mathrm{SPP}}&=&\sqrt{\frac{\epsilon_m\epsilon_d}{\epsilon_m +\epsilon_d}}\label{Eq.spp-index-expression}\\
\lambda_{\mathrm{SPP}}&=&\frac{\lambda_0}{n_{\mathrm{SPP}}}
\end{eqnarray*}}\hfill\parbox{0.1\columnwidth}{\begin{eqnarray}\end{eqnarray}}
where $\lambda_0$ is the free-space incident wavelength, $\epsilon_m,\,\epsilon_d$ are the dielectric constants of the metal and adjacent dielectric respectively, and $n_{\mathrm{SPP}}$ is the effective surface index of refraction.  In the present experiments the dielectric constant of the structured silver sample was measured directly by ellipsometry at $\lambda_0=514.5$~nm and determined to be $\epsilon_m=-9.3+0.18i$.  The dielectric constant of the fused silica substrate is $\epsilon_d=+2.13$, and therefore $\lambda_{\mathrm{SPP}}=309.5\pm 0.1$.

\begin{figure}
  \includegraphics[width=\columnwidth]{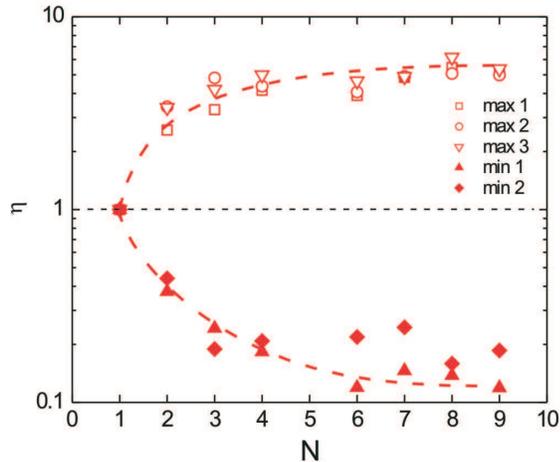}\\
  \caption{(color online) Normalized per-slit transmission intensity maxima and minima vs. the number of slits elements in array $N$.  The labels 1-3 refer to the transmission intensity maxima and minima from left to right shown in Fig.\,\ref{Fig.Intensity-vs-Pitch}}\label{Fig.Intensity-per-slit}
\end{figure}

Figure \ref{Fig.Intensity-per-slit} plots the maximum and minimum values of $\eta$ for each of the $N$ grating series. Enhancement above single-slit transmission up to a factor of $\sim 6$ is observed as $N$ increases up to $N=4$.  At and above $N=4$, adding additional grating elements to the array does not significantly enhance the transmission.  Similar behavior is observed for the transmission minima.  These measurements support the view that transmission enhancement is dominated by nearest-neighbor slit scattering and interference with the strength of the local interaction effectively screening contributions from more distant array elements.  If Bloch surface modes, delocalized over the full extent of the array, played a dominant role, one would expect the per-slit intensity to increase with the number of elements in the array.

\begin{figure}
  \includegraphics[width=\columnwidth]{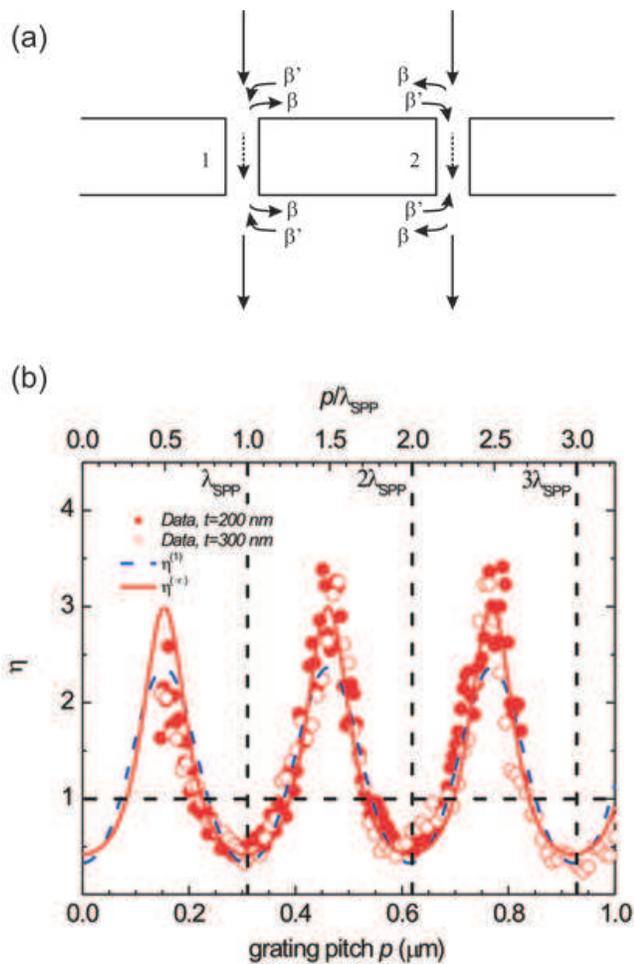}\\
  \caption{(color online) Panel~(a) shows a schematic of the interference model used to fit the two-slit transmission intensity.  Panel~(b) shows the two-slit transmission intensity normalized to single-slit transmission as function of slit-slit separation and plotted on a linear scale.  The data (filled circles) show the transmission profile for a 200~nm thick Ag film; open circles show similar data for a 300~nm thick Ag film.  The dashed line shows a fit using the first-order interference model of Eq.\,\ref{Eq.first-order} (blue curve) and the solid line shows a fit using the infinite-order model of Eq.\,\ref{Eq.inf-order} (red curve).}\label{Fig.Two-slit-FP-Plot}
\end{figure}

Since the positions of transmission suppression and enhancement as a function of period are essentially independent of $N$, we can analyze the mechanism responsible for modulation by concentrating on the simplest case $N=2$.  The normalized per-slit transmission intensity $\eta$ of an array of slit pairs with varying pitch $p$ is shown in Fig.~\ref{Fig.Two-slit-FP-Plot}b.  The intensity $\eta$ is plotted on a linear scale as a function of $p$ for devices milled into a Ag film of thickness $t=300$~nm (hollow circles) in addition to the Ag film of thickness $t=200$~nm described earlier (solid circles, replotted from Fig.\,\ref{Fig.Intensity-per-slit}).  Comparison of the two data sets shows that the $\eta(p)$ modulation is essentially invariant with $t$ and is therefore governed by the interaction between the two slits mediated by surface waves running along both facets of the structured metal film. Periodic minima are measured at slit-slit distances corresponding to integer multiples of $\lambda_{\mathrm{SPP}},\,(p=n\lambda_{\mathrm{SPP}}, n=1,2,\ldots)$.  This observation is consistent with recent theoretical predictions for a two-slit system\,\cite{JUH06} (albeit for a structure with only one SPP-sustaining surface).

We have developed a simple, first-order model for $\eta(p)$.  Figure \ref{Fig.Two-slit-FP-Plot}a shows the essential idea.  Two surface waves, one at each slit, are generated by diffractive scattering of a normally incident plane wave.  The respective SPPs counterpropagate along the surface, creating a standing surface wave and interfering with the incident wave at the opposite slit.  The intensity transmitted to the other side of the metal film is proportional to the resulting modulated intensity at each slit opening.  An identical process takes place on the exit side of the film.  Counterpropagating surface waves are launched by diffractive scattering at the slit exits and interfere with the directly propagating mode at the opposite slit location.  In Fig.\,\ref{Fig.Two-slit-FP-Plot}a\, $\beta$ designates the amplitude launching efficiency of the SPP by a given slit at either entrance or exit aperture.  More specifically $\beta$ is defined by the ratio of the launched SPP magnetic H-field complex amplitude to that of the incident wave at the slit entrance or to that of the propagated mode at the slit exit. Similarly $\beta^{\prime}$ designates the conversion efficiency of the SPP complex amplitude back to propagating modes at either side of the slit opening.  This first-order interference process yields a net transmission intensity (normalized to that of an isolated slit) given by,
\begin{equation}
\eta^{(1)}(p)=\left\{1+\left(\beta_0\beta^{\prime}_0\right)^2+2\beta_0\beta_0^{\prime}\cos\left[\left(\frac{2\pi}{\lambda_\mathrm{surf}}\right)p+\varphi\right]\right\}^2
\label{Eq.first-order}
\end{equation}
In Eq.\,\ref{Eq.first-order} $\lambda_{\mathrm{surf}}$ is the surface wavelength, $\beta_0=|\beta|, \beta_0^{\prime}=|\beta^{\prime}|$ are the magnitudes of the launch and conversion efficiencies at the slits.  The phase $\varphi=\arg\left(\beta\beta^{\prime}\right)$ is the phase associated with the SPP\,$\leftrightarrow$\,propagating wave conversion, exclusive of the phase accumulated along the surface, $(2\pi/\lambda_{\mathrm{surf}})p$.  Refining the model by taking multiple reflections into account at the slits results in the following closed-form expression,
\begin{equation}
\eta^{(\infty)}(p)=\left\{1+\left(\beta_0\beta^{\prime}_0\right)^2-2\beta_0\beta_0^{\prime}\cos\left[\left(\frac{2\pi}{\lambda_\mathrm{surf}}\right)p+\varphi\right]\right\}^{-2}
\label{Eq.inf-order}
\end{equation}
A fit of $\eta^{\infty}(p)$ to the combined set of experimental transmission data for both $t=300$~nm and $t=200$~nm is shown in Fig.\,\ref{Fig.Two-slit-FP-Plot}b (solid red curve).  Excellent agreement is obtained using fitting parameters $\lambda_{\mathrm{surf}}=307$~nm, $\beta_0\beta_0^{\prime}=0.24$, and $\varphi=\pi$.  The best-fit value for $\lambda_{\mathrm{surf}}$ is slightly less than the theoretical value $\lambda_{\mathrm{SPP}}=309.5$~nm.  Considering that most of the surface wave propagation takes place in the ``near zone"\,\cite{GAV06a,KGA07,GAW07} where transient modes contribute to form a composite surface wave, this result is not surprising.  The shape of $\eta^{\infty}(p)$ is reminiscent of the transmission characteristics of a ``lossy" two-mirror Fabry-P\'erot resonator of free spectral range $\Delta\lambda=\lambda_{\mathrm{surf}}$, full-width at half-maximum $\partial\lambda=0.29\lambda_{\mathrm{surf}}$ and finesse $F=\Delta\lambda/\partial\lambda=3.4$.  The Fabry-P\'erot profile suggests the influence of multiple surface wave reflections at the slit sites.   A plot of the first-order model $\eta^{(1)}(p)$ is also included in Fig.\,\ref{Fig.Two-slit-FP-Plot}b (dashed blue curve), using the fitting parameters above. The essential profile of the normalized transmission as a function of $p$ is already well reproduced by $\eta^{(1)}(p)$.  The first-order fit suggests that the formation of transmission minima is predominantly controlled by interference at the slit openings rather than by the presence of higher order multiple reflections.  The positioning of the minima at $p=n\lambda_{\mathrm{SPP}},\, (n=1,2,\ldots)$ is due to $\varphi=\pi$ and is in agreement with the findings of Ref.\,\cite{JUH06} although the search for a simple but convincing physical explanation for this phase continues.

In summary we have measured the transmitted far-field intensity through a series of subwavelength slit arrays as a function of array pitch and have determined that the minimum per-slit transmission at the array output facet occurs for an array pitch equal to the wavelength of the surface plasmon polariton.  We have also determined that the per-slit transmitted intensity does not increase appreciably above an array size greater than $N=4$.

Support from the Caltech Kavli Nanoscience Institute, the National Science Foundation under Grant DMR 0606472 and the use of facilities at the Center for Science and Engineering of Materials, an NSF Materials Research Science and Engineering Center at Caltech, is gratefully acknowledged.
Support from the Minist{\`e}re d{\'e}l{\'e}gu{\'e} {\`a} l'Enseignement sup{\'e}rieur et {\`a} la Recherche under the programme
ACI-``Nanosciences-Nanotechnologies," the R{\'e}gion Midi-Pyr{\'e}n{\'e}es
[SFC/CR 02/22], and FASTNet [HPRN-CT-2002-00304]\,EU Research
Training Network, as well as from the research foundation FAPESP of the State of S\~ao Paulo, Brazil is also gratefully acknowledged.

\end{document}